\documentclass[conference]{IEEEtran}
\IEEEoverridecommandlockouts
\usepackage{cite}
\usepackage{amsmath,amssymb,amsfonts}
\usepackage{algorithmic}
\usepackage{graphicx}
\usepackage{textcomp}
\usepackage{xcolor}

\usepackage{multirow}

\newcommand*{\fig}[4]{ 
\begin{figure}[ht]
\centerline{\includegraphics[width=#4\linewidth]{#1}}
\caption{#2 }
\label{#3}
\end{figure}

}

\newcommand*{\crc}[0]{\textit{code review comment}}
\newcommand*{\crcs}[0]{\textit{code review comments}}

\newcommand*{\Crcs}[0]{\textit{Code review comments}}

\newcommand*{\figref}[1]{Fig.~\ref{#1}}

\makeatletter
\newcommand{\getAuthors}[1]{\begingroup\edef\x{\endgroup\noexpand\expandafter\@gobbletwo\noexpand#1}\x}
\makeatother

\newcommand*{\citesrf}[1]{
\getAuthors{\cite{#1}}~\cite{#1}
}

\newcommand*{\citep}[1]{\cite{#1}}
\newcommand*{\citet}[1]{\citesrf{#1}}

\def\BibTeX{{\rm B\kern-.05em{\sc i\kern-.025em b}\kern-.08em
    T\kern-.1667em\lower.7ex\hbox{E}\kern-.125emX}}
\begin{document}

\title{
Exploring the Advances in Identifying Useful Code Review Comments
}

 
\IEEEpubid{\begin{minipage}{\textwidth}\ \\[12pt]
\\
  \emph{This paper has been accepted for inclusion in the Proceedings of \\ the 17th ACM/IEEE International Symposium on Empirical Software\\ Engineering and Measurement (ESEIW/ESEM 2023)}
\end{minipage}} 
\author{\IEEEauthorblockN{Sharif Ahmed}
\IEEEauthorblockA{\textit{Computer Science Department} \\
\textit{Boise State University}\\
Boise, ID, USA \\
sharifahmed@u.boisestate.edu}
\and
\IEEEauthorblockN{Nasir U. Eisty}
\IEEEauthorblockA{\textit{Computer Science Department} \\
\textit{Boise State University}\\
Boise, ID, USA \\
nasireisty@boisestate.edu}
}

\maketitle

\begin{abstract}
Effective peer code review in collaborative software development necessitates useful reviewer comments and supportive automated tools. 
\Crcs\ are a central component of the Modern Code Review process in the industry and open-source development. 
Therefore, it is important to ensure these comments serve their purposes. 
This paper reflects the evolution of research on the usefulness of \crcs. 
It examines papers that define the usefulness of code review comments, mine and annotate datasets, study developers' perceptions, analyze factors from different aspects, and use machine learning classifiers to automatically predict the usefulness of \crcs. 
Finally, it discusses the open problems and challenges in recognizing useful \crcs\ for future research.    

\end{abstract}


\section{Introduction}

Software engineers have been using code review for decades to significantly improve code quality, enhance collaboration, facilitate knowledge transfer, enforce coding standards, and save time and cost.
Initially, software developers and testers identified software errors and defects in person through a process called \textit{Formal Code Review} or \textit{Fagan Inspection}~\citep{fagan2002design}. 
This process involved six phases: planning, overview, preparation, inspection meeting, rework, and follow-up.
However, with the advancement of the Globalization, software development teams have become more distributed globally, resulting in a decline in the use of this in-person source-code quality inspection process. 
In addition, the appearance of online communication platforms facilitated developers to conduct such discussions asynchronously and informally.
This online tool-based source code inspection technique is known as \textit{Modern Code Review (MCR)}.

The MCR 
process has been adopted by both industry and open-source developers. 
However, organizations may customize the process to fit their specific needs, using different tools like GitHub or Gerrit and implementing various policies and working cultures.
Generally, the MCR process involves five phases: review request, reviewer selection, review task selection, code review/checking, and review feedback. 
Reviewers can provide feedback using two approaches. 
The first is through an annotation on IDE, which is similar to reviewing a paper document with a pen or pencil, as seen in the Rich Code Annotation tool~\cite{priest2006rca}. 
The second approach is through written feedback in natural language, emoji, emoticons, animation, and votes
, which is known as \crc.

\begin{figure*}[tbp]
\centerline{\includegraphics[trim={0 0cm 0 0.0cm},clip, width=0.99\linewidth]{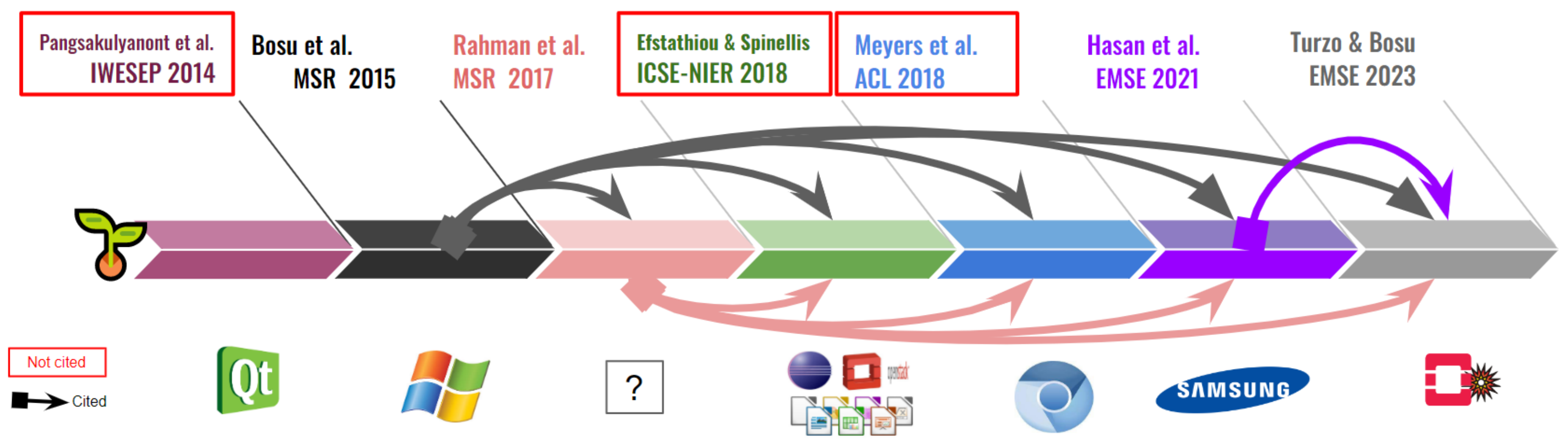}}

\caption{Progression of Works on Usefulness of \crcs}
\label{fig_timeline}
\end{figure*}

The practice of code review aims to produce quality software while saving time and money. 
However, the intermediate outcome of the code review process, \crc, has the most significant impact on regulating the outcomes. 
The importance of \crcs\ was demonstrated in Kononenko et al.'s empirical study of the Mozilla code review process~\cite{kononenko2016code}. 
However, Bosu et al.~\cite{bosu2015characteristics} found that 34.5\% of the \crcs\ were not useful at Microsoft. 
Previous studies have defined the usefulness of \crcs~\citep{pangsakulyanont2014assessing, bosu2015characteristics, meyers2018dataset}, 
mined and annotated datasets of \textbf{useful} 
\crcs~\citep{rahman2017predicting, meyers2018dataset,turzo2023makes}, 
studied developers' perceptions in commercial~\cite{bosu2015characteristics} and open-source projects~\cite{turzo2023makes}, 
analyzed factors from various aspects~\citep{pangsakulyanont2014assessing,bosu2015characteristics,rahman2017predicting, meyers2018dataset, efstathiou2018code, hasan2021usingCRA,turzo2023makes}, 
adapted fine-grained taxonomy~\cite{turzo2023makes} for useful \crcs, and 
used machine learning classifiers to automatically predict the usefulness of \crcs~~\citep{pangsakulyanont2014assessing,bosu2015characteristics,rahman2017predicting, meyers2018dataset,hasan2021usingCRA}. 
Although these studies have progressed the research problem since 2014 (\figref{fig_timeline}), some works did not reference their previous related papers. 
This observation has prompted us to reflect retrospectively on the literature on the usefulness of \crcs.

To reflect on these studies, we conducted a literature review, focusing on studies explicitly or implicitly exploring the usefulness of \crcs. 
Next, we delved deeper into the studies and evaluated their contributions, approaches, limitations, and takeaways in Sec-II. 
Our reflection paper is intended to assist researchers and developers in advancing ongoing research or identifying an appropriate strategy for analyzing or identifying the usefulness of \crcs.

 
 
 


\section{Reflection}
In this section, we reflect on previous research on predicting and analyzing the usefulness of \crcs.
We selected a set of studies and refer to these papers as `key papers' (\figref{fig_timeline}).

\textit{\textbf{1. Key Paper Selection:} }
We utilized a multi-step approach to identify the key papers for our study. Initially, we searched Google Scholar using the query string ``useful code review comments." Subsequently, we performed backward and forward snowballing to expand our search further.
Regarding inclusion criteria, we considered papers published in peer-reviewed workshops, conferences, or journals. We excluded pre-print papers and magazine articles from our selection process.
While many papers explore the efficiency and value of the Modern Code Review (MCR) process, including its various phases, techniques, recommendations, and generation, our focus was explicitly on papers that addressed or analyzed the usefulness of English code review comments. Consequently, we excluded papers that mentioned code review comments without explicitly delving into assessing or identifying their usefulness.
As a result of our rigorous selection process, we identified seven studies that met our criteria, and we presented a synthesis of the extracted data in Table~\ref{tbl_matrix}.

\textbf{\textit{2. Exploration:}} 
Here, we explore different aspects (\textit{
A-J
}) of code review comments' usefulness 
from usefulness definition to automatic classification and its evaluation.

 \begin{figure}[hb]
\fbox{\includegraphics[trim={0 4.5cm 0 3.5cm},clip, width=0.99\linewidth]{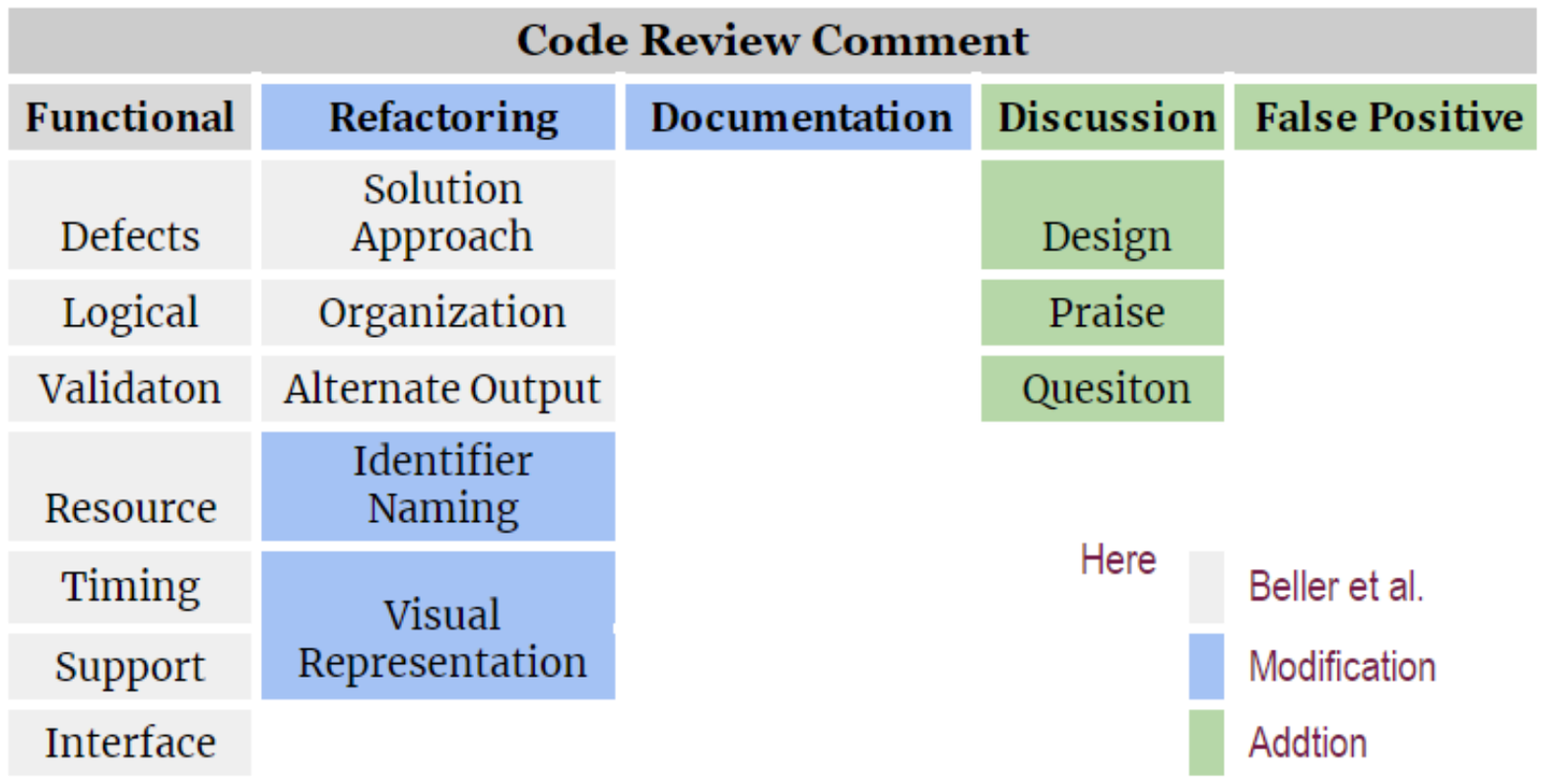}}

\caption{ Taxonomy of  \Crcs\  by Turzo \& Bosu~\cite{turzo2023makes}} 
\label{taxonomy}

\end{figure}

 \begin{table*}[htbp]
\centering
\small
\caption{Synthesis of Existing Works Regarding Usefulness of Code Review Comments}
\label{tbl_matrix}
\raisebox{-\totalheight}{\includegraphics[trim={0 2cm 0cm 0cm},clip, width=\linewidth]{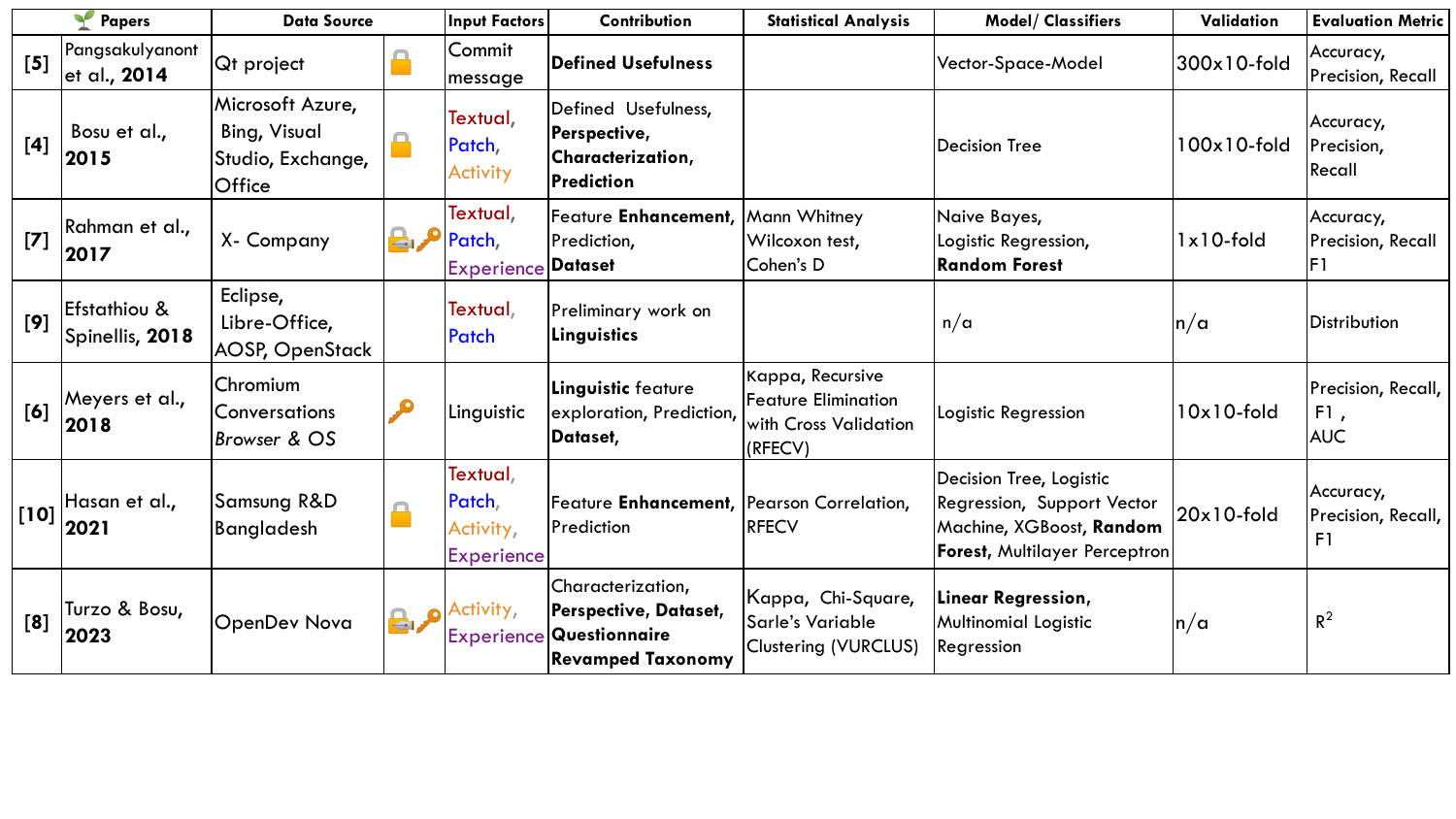}}
\end{table*}

\begin{figure*}[htbp]

\centerline{\includegraphics[trim={0 0cm 0 0.0cm},clip, width=0.99\linewidth]{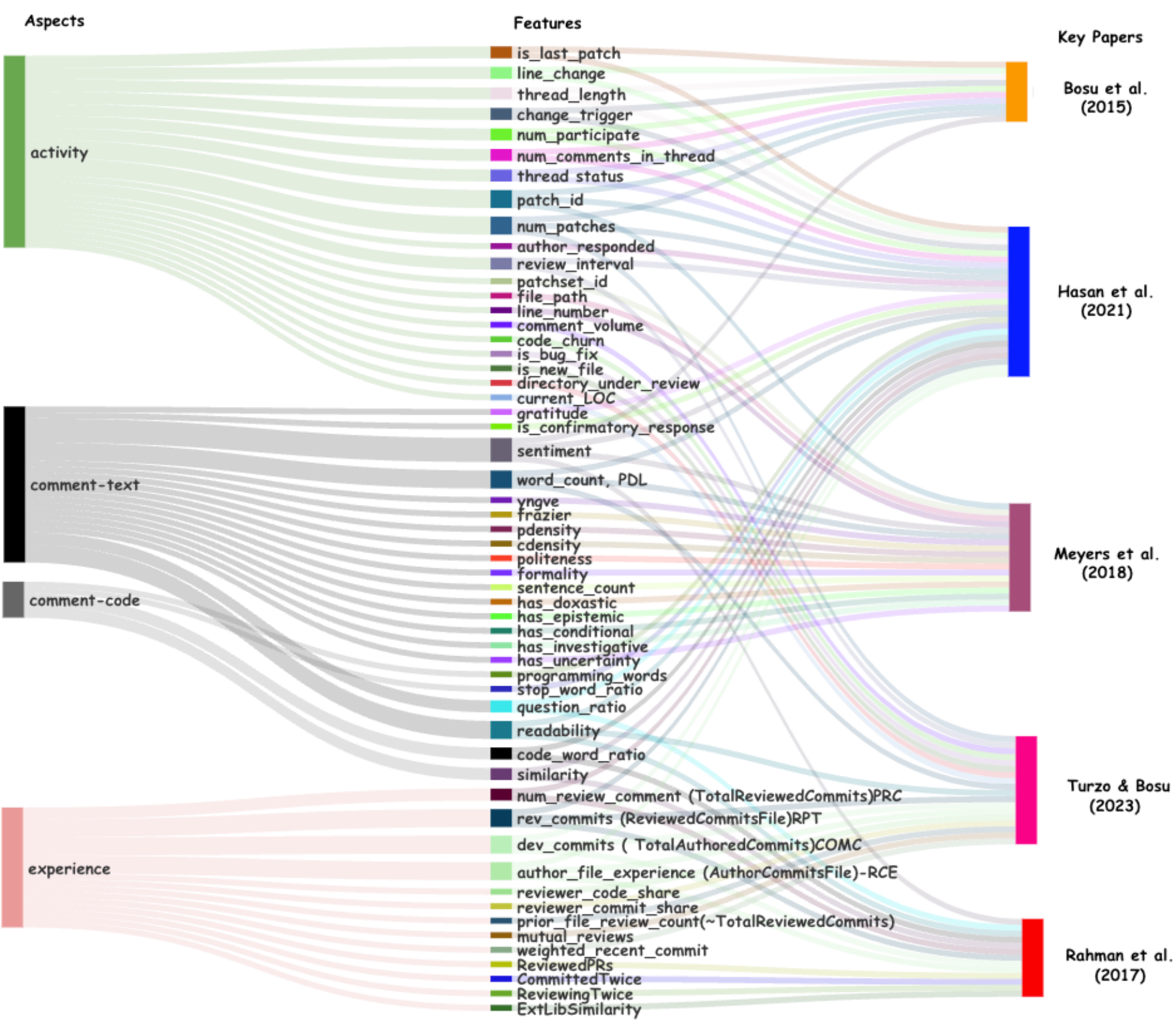}}

\caption{An examination of the features employed to date for predicting/ analyzing the usefulness of \textit{Code Review comments}. }
\label{fig_fts_dg}
\end{figure*}


\subsection{Usefulness Definition }
In 2014, Pangsakulyanont et al.~\cite{pangsakulyanont2014assessing} established a definition for \crcs\, categorizing them as either {\bf useful} or {\bf useless} according to the degree of similarity between the commit message of the author and the reviewer's \crc\ for a given code change. 
In cases where the usefulness of the \crc\ was unclear, they classified it as \textbf{undetermined}. 

In the following year, Bosu et al.~\cite{bosu2015characteristics} conducted a study at Microsoft, in which seven developers classified \crcs\ as \textbf{useful, somewhat useful,} and \textbf{not useful}. 
While building a usefulness prediction model, they treated \textbf{somewhat useful} as \textbf{useful}. 
The authors defined a \crc\ as \textbf{useful} if it led to a code change in nearby source-code lines. 
Their experiment investigated the proximity between the code-change and the \crc, with nearness ranging from 1-10 lines.  
They discovered that changes one line away had the lowest rates of \textit{false positives} and \textit{false negatives} for usefulness classification.

Recently, in 2023, Turzo \& Bosu \cite{turzo2023makes} have defined and annotated comments as \textbf{useful} if the author explicitly acknowledges the reviewer's identified issue as good, implicitly acknowledges the feedback by implementing recommended changes, or explicitly retains the feedback for future modifications.  

\subsection{Taxonomy of Code Review Comments \& Usefulness}
Turzo \& Bosu~\cite{turzo2023makes} expanded on the labeling of their dataset of \crcs\ beyond the binary \textbf{useful} and \textbf{not-useful} labels. 
They introduced 18 categories to gain a deeper understanding of the usefulness of \crcs. 
These categories were adapted from Beller et al.'s~\cite{beller2014modern} taxonomy of \crcs\ for changes.
We have redrawn their taxonomy 
in \figref{taxonomy}. 

\subsection{Understanding Developers' Perception}
Researchers have investigated how developers perceive the usefulness of \crcs\ in both commercial~\cite{bosu2015characteristics} and open-source~\cite{turzo2023makes} communities. 
We present a concise overview of their findings below.

\subsubsection{Microsoft Study (2015)}
Bosu et al.~\cite{bosu2015characteristics} conducted a three-stage experiment at Microsoft to characterize the usefulness of \crcs. 
Firstly, they interviewed developers to gain insight into useful \crcs. 
Secondly, they annotated the usefulness based on developers' understanding and built a usefulness predictor. 
Thirdly, they conducted an empirical analysis of the identified factors. 
Based on prior studies, the authors focused on two types of factors: (i) the \textit{reviewers} and (ii) the \textit{changeset}. 
The study suggests that expert reviewer selection is crucial, but the inclusion of novice reviewers can also be beneficial for disseminating expertise. 
The results also indicate that review effectiveness decreases as the number of files in the patches increases. 

\subsubsection{ OpenDev Study (2023)}
Turzo \& Bosu~\cite{turzo2023makes} conducted a three-stage experiment to gain a deeper understanding of the usefulness of \crcs. 
Firstly, they manually categorized and labeled the usefulness of \crcs. 
Secondly, they obtained insight from OSS developers through an online survey. 
Thirdly, they performed a regression analysis on contextual and participant factors with the usefulness of \crcs. 
Kononenko et al.~\cite{kononenko2015investigating} performed a similar analysis for buggy changes at \textit{Mozilla}. 
Turzo \& Bosu obtained 160 usable empirical responses from OpenDev developers via email while maintaining the IRB protocol. 
The OpenDev developers suggested that usefulness depends on verbal and process aspects in addition to technical contributions. 
They also emphasized that functional \crcs\ matter the most and hold diverse views on the usefulness, including praise, documentation, design discussion, resource synchronization, and visual representation.

\subsection{Factors and Features}

Initially, Pangsakulyanont et al.~\cite{pangsakulyanont2014assessing} employed the Vector Space Model with cosine similarity measure between the commit message and the \crc\ on a code-change to classify usefulness. 

Subsequently, in 2015, Bosu et al.~\cite{bosu2015characteristics} identified the factors of useful \crcs\ through an empirical study at Microsoft and defined ``usefulness''. 
They also developed a Decision Tree Classifier to model a usefulness classifier. 
To generate features for the classifier, they utilized textual properties of \crcs\ and attributes of review activities
(\figref{fig_fts_dg}). 

In another study, Kononenko et al.~\cite{kononenko2015investigating} conducted an empirical investigation on the open-source project \textit{Mozilla}. 
They discovered that 54\% of the reviewed changes failed to detect bugs in the code. 
They also introduced factors, such as code author and reviewer experience, that impact the quality of code reviews.
Rahman et al.~\cite{rahman2017predicting} collected \crcs\ from commercial projects and developed a model to predict their usefulness. 
Their approach involved using textual properties of the \crcs\ 
and features related to the developers' experience~(\figref{fig_fts_dg}). 
However, unlike Bosu et al.~\cite{bosu2015characteristics}, they did not incorporate any features related to the code review activities.

Previous studies, such as~\cite{bosu2015characteristics,rahman2017predicting,hasan2021usingCRA}, have achieved better results by including textual features and developers' experience and review activity features. 
However, in projects with no prior review or record of reviewers and developers, these important features will be missing in these studies. 
Efstathiou and Spinellis~\cite{efstathiou2018code} proposed measuring usefulness using linguistic semantics to address this issue. 
Concurrently, Meyers et al.~\cite{meyers2018dataset} used several linguistic features, including ``text complexity," ``density," ``formality," ``politeness," ``sentiment," and ``uncertainty," to classify \crcs. 
Recently, OSS developers also stated politeness and comprehension as key factors~\cite{turzo2023makes}.

Hasan et al.~\cite{hasan2021usingCRA} developed an in-house web-based application to reward reviewers at \textit{Samsung Research Bangladesh} for their useful feedback. 
They did not consider any linguistic features from Meyers et al.~\cite{meyers2018dataset}.
The authors used features from three aspects of prior works: textual, experience, and review activities to predict the usefulness of reviewers' \crcs. 
Additionally, they extended the existing features by incorporating features related to code review activities 
(\figref{fig_fts_dg}). 
They also introduced a new textual feature called word count to identify short comments.
Furthermore, they presented several features for predicting the usefulness of \crcs\ from the literature and their work, but these features are limited to their model.

We have compiled all features and categorized them based on their aspects, such as review context or activity, developers' experience, and textual features from \crcs. 
Additionally, we have divided the comment-textual features into \textit{comment-text} and \textit{comment-code} aspects. 
The features mapped from various aspects to key papers are presented in \figref{fig_fts_dg}. 
For instance, \figref{fig_fts_dg} shows that the `change\_trigger' feature used by Bosu et al.~\cite{bosu2015characteristics} and 
Hasan et al.~\cite{hasan2021usingCRA} was derived from the code review `activity' aspect.

\subsection{Available Data Sources}
The dataset by Bosu et al.~\cite{bosu2015characteristics} and Hasan et al.~\cite{hasan2021usingCRA} cannot be made public due to the Non-Disclosure Agreements (NDAs) they have with Microsoft and Samsung R\&D Bangladesh, respectively. 
However, we have identified three other datasets relevant to the usefulness of \crcs. 
The statistics of usefulness labels for all available datasets are presented in \figref{fig_dataset_stats}.

\subsubsection{RevHelper}
In 2017, Rahman et al.~\cite{rahman2017predicting} used Github API to collect \crcs\ from four commercial subject systems of an anonymous company. 
They manually annotated each \crc\ as either \textbf{useful} or \textbf{non-useful}, based on the explanation of usefulness provided in the prior work by Bosu et al.~\cite{bosu2015characteristics}. 
The dataset includes 
879 \textbf{useful} and 602 \textbf{not-useful} comments. 
The authors distributed the dataset into two sets: a \textit{train-test set}
and a \textit{validation set}.

\subsubsection{Chromium Conversation}
In 2018, Meyers et al.~\cite{meyers2018dataset} developed a dataset from code review in Google Chromium Project.
The open-source code review tool used in Chromium is known as \textit{Rietveld}. 
Using RESTful API, the authors obtained 2,855,018 publicly accessible code review comments from 2008 to 2016. 
They then selected comments posted by reviewers. 
Next, they automatically identified acted-upon \crcs\ by exploiting Rietveld's ``Done" click feature. 
Their annotation principle is similar to the RevHelper \cite{rahman2017predicting}.
The dataset consists of 2994 \textbf{acted-upon} (i.e., \textbf{useful}) and 800 \textbf{not-}(known to be)-\textbf{acted upon} \crcs.

 \subsubsection{OpenDev Comment} 
In 2023, Turzo \& Bosu~\cite{turzo2023makes} developed a dataset from OpenDev's Nova OSS project. 
They utilized the Gerrit Miner tool to mine 795,226 publicly available code-review records from 2011 to 2022. 
The resulting dataset consists of 2,052 \textbf{useful} and 602 \textbf{not-useful} \crcs\ that were manually annotated from 18 categories and 5 comment groups. \\
 


\textit{Annotation Verification:}
The authors of the dataset papers also verified their annotations differently.
Rahman et al. 
manually annotated the RevHelper~\cite{rahman2017predicting} dataset and conducted a random cross-check. 
Meyers et al.~\cite{meyers2018dataset} automatically annotated \textbf{acted-upon} \crcs\ and manually verified 23\% of the sample size, resulting in an inter-annotator agreement score of 0.89. 
They then manually inspected the 2047 \textbf{not acted-upon} \crcs\ that were automatically considered and included 800 correctly identified \textbf{not acted-upon} \crcs\ in their final dataset. 
Turzo and Bosu~\cite{turzo2023makes} reported inter-rater reliability scores of 0.68 and 0.84 for their manual \textit{comment-category} and \textit{comment-usefulness} annotation, respectively.


\subsection{Feature Analysis and Selection}
Rahman et al.~\cite{rahman2017predicting} compared the effectiveness of input features with that of \crcs\ in predicting the usefulness of the latter. 
They utilized the \textit{Mann Whitney Wilcoxon test} to identify differences in features between useful and non-useful \crcs, and \textit{Cohen's D} as an effect-size measure. 
In contrast, Hasan et al.~\cite{hasan2021usingCRA} used the \textit{Pearson Correlation} measure to interpret their input features. 
 %
In their study, Rahman et al.~\cite{rahman2017predicting} compared the textual characteristics of useful and non-useful comments. 
They found that the \textit{code element ratio, stop word ratio}, and \textit{conceptual similarity} were statistically significant, while \textit{reading ease} and \textit{question ratio} were not. Additionally, they explored the relationship between developers' experience and useful \crcs. They found that \textit{code ownership} and \textit{reviewership} were statistically significant, while the relationship between developers' experience and usefulness was not straightforward, which is consistent with Bosu et al.~\cite{bosu2015characteristics}.
\fig{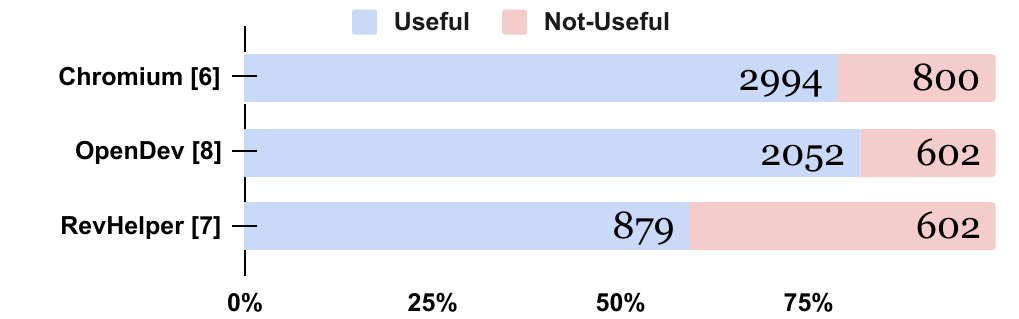}{Available Datasets \cite{rahman2017predicting,meyers2018dataset,turzo2023makes}}{fig_dataset_stats}{0.95}
In 2018, Efstathiou and Spinellis~\cite{efstathiou2018code} conducted a preliminary examination using the textual data of review messages obtained 
from four open-source projects (Table-\ref{tbl_matrix}).
Their analysis revealed a possible avenue for exploring linguistic semantics to identify beneficial \crcs; however, verifying the applicability would require the input of skilled linguists.

Meyers et al.~\cite{meyers2018dataset} used Recursive Feature Elimination with Cross-Validation (RFECV) to filter out the less contributing features. Hasan et al.~\cite{hasan2021usingCRA} also used RFECV after excluding highly correlated features.

Turzo \& Bosu~\cite{turzo2023makes} initially applied Sarle's Variable Clustering (VURCLUS) technique to identify highly correlated features. 
They then used Spearman's correlational hierarchical clustering approach to remove features with a high correlation coefficient of 0.7+ from their regression model.  
They also used the Chi-Square test to check the relationship between respondents' demographics and responses.  

\subsection{Models} 
Three key papers employed only a single algorithm to classify useful \crcs.
Specifically, Pangsakulyanont et al.~\cite{pangsakulyanont2014assessing} used the Vector Space Model to measure similarity, Bosu et al.~\cite{bosu2015characteristics} employed the Decision Tree Classifier, and Meyers et al.~\cite{meyers2018dataset} utilized the Logistic Regression Classifier. 
Other studies explored multiple classification algorithms, such as the Decision Tree Classifier~\cite{bosu2015characteristics, hasan2021usingCRA}, Naive Bayes Classifier~\cite{rahman2017predicting}, Logistic Regression Classifier~\cite{rahman2017predicting, hasan2021usingCRA, meyers2018dataset}, Random Forest Classifier~\cite{rahman2017predicting,hasan2021usingCRA}, Support Vector Machine~\cite{hasan2021usingCRA}, Multilayer Perceptron Classifier~\cite{hasan2021usingCRA}, and eXtreme Gradient Boosting Classifier~\cite{hasan2021usingCRA}. 
Two studies~\cite{rahman2017predicting,hasan2021usingCRA} found Random Forest Classifier as the best performing model in their evaluation.
For regression analysis, Turzo \& Bosu~\cite{turzo2023makes} used Linear Regression and Multinomial Logistic Regression models.

All previous models had imbalanced training datasets 
except for Hasan et al.~\cite{hasan2021usingCRA} 
who employed the Synthetic Minority Over-sampling Technique (SMOTE).

\subsection{Evaluation Metrics}

TABLE~\ref{tbl_matrix} indicates that the majority of key papers evaluated the effectiveness of their code review comment usefulness prediction using the {\bf accuracy, precision, recall,} and {\bf F1-measure} metrics. 
One of the papers also reported using the {\bf AUC} (Area Under the Curve) metric as an evaluation method. 
Another study utilized $\pmb{R^2}$ to measure the efficacy of their regression models.

Though they used different datasets and evaluation measures,
the \textbf{accuracy} and \textbf{F1$_{w.avg.}$} found in the key papers range from \~{}63 to \~{}87\%.
We also observed that they could achieve \~{}6 to \~{}10\% higher accuracy than the majority class (i.e., \% useful comments). 






\subsection{Model Comparison}
Rahman et al.~\cite{rahman2017predicting} conducted a comparison with several simulated variants of their previous work \cite{bosu2015characteristics}.
Similarly, Hasan et al.~\cite{hasan2021usingCRA} replicated two of their earlier models~\cite{bosu2015characteristics, rahman2017predicting} with slight modifications and compared their performance.
Their replication addressed the issue of inaccessible features such as {\it external library, keyword ratio, reading ease} in RevHelper~\cite{rahman2017predicting}. 
However, they also obtained some features by replacing new tools with the original ones. 
For example, while Bosu et al.~\cite{bosu2015characteristics} used a text-sentiment analyzer, Hasan et al.~\cite{hasan2021usingCRA} replaced it with a code-review sentiment analyzer in their replication. 

Various techniques, including cross-validation, offer a way to mitigate the risk of over-fitting or selection bias in model development.
These techniques also provide valuable insights into the generalizability of a model, particularly its ability to perform accurately on an independent dataset.
All of the prediction works~\cite{pangsakulyanont2014assessing},~\cite{bosu2015characteristics},~\cite{ rahman2017predicting},~\cite{ meyers2018dataset}, and ~\cite{ hasan2021usingCRA} evaluated their models using 10-fold cross-validation, with 300, 100, 0, 10, and 20 repetitions, respectively.

 \subsection{Limitations}
The literature exhibits two major problems.

\subsubsection{Data Unavailability or Limited Availability}
Bosu et al.~\cite{bosu2015characteristics} and Hasan et al.~\cite{hasan2021usingCRA} could not share their dataset for proprietary restrictions.
Due to the same reason, Rahman et al.~\cite{rahman2017predicting} shared a partial dataset, and the code changes were excluded. 
However, the other two accessible~\cite{meyers2018dataset,turzo2023makes} datasets have identifiers instead of direct code changes. 
Unfortunately, due to time constraints, it was not feasible to assess the viability of retrieving the code changes from the given linking identifiers.
However, the existing usefulness prediction models~\citep{bosu2015characteristics, rahman2017predicting, hasan2021usingCRA} demand code-changes (aka patches)  to classify usefulness which is absent in available datasets.  

\subsubsection{Review Activity and Experience Records}
The existing usefulness prediction models~\citep{bosu2015characteristics, rahman2017predicting, hasan2021usingCRA} require review activity, developer experiences, or both, which may not be available for a new developer, project, or company. 


\section{ Discussion }

Researchers have attempted to define the usefulness of code review comments, study developers' perceptions, analyze underlying factors, extract features from these factors, build datasets, and classify the usefulness of code review comments. 
However, due to proprietary restrictions, most studies have not made their experimental datasets available
~\cite{bosu2015characteristics,hasan2021usingCRA,rahman2017predicting}
. 
These unavailable data limit the replication process and public reviews.

While examining existing features, we considered \textit{comment-code} as a new aspect. 
New potential features can be adopted under this aspect to predict the usefulness of \crcs.
Crucially, a key paper suggested the linguistic approach to identify the usefulness of \crcs~\cite{efstathiou2018code}. 
Cutting-edge NLP tools can be applied accordingly.
Furthermore, researchers have evaluated the performance of their models using various metrics. 
Based on the reported performance, it is open that there is still room for improvement in the existing models. 

Some works were not mentioned in their subsequent papers (\figref{fig_timeline}), and several reasons could exist. 
One possible explanation is that two papers were published in the same year. 
Another reason could be using different category names for usefulness ~\cite{meyers2018dataset} 
or publication in another venue. 


Two empirical studies have corroborated many of the findings from other studies, both empirical and non-empirical. 
These findings include the importance of factors such as reviewer experience, text-readability, timeliness, sentiment, and reviewers' tenure, as well as the inverse relationship between the usefulness of \crcs\ and the size of the changes being reviewed. 
 MacLeod et al.~\cite{macleod2017code} reported challenges and best practices for the MCR process. 
They found factors of \crcs\ such as \textit{comprehension},\textit{ being respectful} affect the MCR process, and these are relatable to the features adopted in the key papers\cite{rahman2017predicting,bosu2015characteristics} such as  readability, sentiment, 
etc.

The findings regarding developer experience and its relationship with the usefulness of \crcs\ have been more mixed, with the Microsoft study finding a positive association between reviewers' tenure and usefulness. In contrast, the OpenDev study found a negative association with project tenure. 
To better understand these discrepancies or discover new factors, it may be useful for researchers to conduct qualitative studies with developers and to perform further data mining analyses. 

\section{Conclusion}
This paper reflects existing literature and outlines the research trajectory on the usefulness of \crcs. It identifies ten 
areas, including defining the usefulness of \crcs, adapting taxonomies, exploring developer perceptions, extracting factors and features, data mining and annotation, conducting feature analysis, predicting usefulness, evaluating approaches, and major limitations. 
Moving forward, developers may consider creating a tool that can identify not-useful \crcs\ in advance to prevent before commenting.
Other related tasks, such as automatic code review comment generation, reviewer or developer recommendation, static analysis, etc., involve \crcs. 
These tasks can leverage the useful \crcs\ prediction models and enhance the effectiveness of the respective tasks or the entire code review process. 

\IEEEtriggeratref{8}

\bibliographystyle{IEEEtran}%
\bibliography{references}

\end{document}